\documentstyle[aps,prl,twocolumn,epsf,floats]{revtex}

\newcommand{\beq}{\begin{equation}}
\newcommand{\eeq}{\end{equation}}
\newcommand{\bea}{\begin{eqnarray}}
\newcommand{\eea}{\end{eqnarray}}
\newcommand{\eps}{\epsilon}

\newcommand{\nn}{\nonumber}
\newcommand{\benn}{\begin{displaymath}}
\newcommand{\eenn}{\end{displaymath}}
\begin{document}

\title{\bf \Large
Energy Density Functional Approach to Superfluid Nuclei
}
\vspace{0.75cm}
\author{Yongle Yu and Aurel Bulgac}
\vspace{0.50cm}
\address{Department of Physics, University of
Washington, Seattle, WA 98195--1560, USA}
\maketitle

%\today

\begin{abstract}

We show that within the framework of a simple local nuclear energy
density functional (EDF), one can describe accurately the one-- and
two--nucleon separation energies of semi--magic nuclei. While for the
normal part of the EDF we use previously suggested parameterizations,
for the superfluid part of the EDF we use the simplest possible local
form compatible with known nuclear symmetries.

\end{abstract}

\draft

% 21.60.Jz Hartree-Fock and random phase approximations
% 21.60.-n Nuclear-structure models and methods
% 21.30.Fe Forces in hadronic systems and effective interactions

A steady transition is taking place during the last several years
from the mean--field description of nuclear properties in terms of
effective forces to an energy density functional approach (EDF). A
significant role is played in this transition process by the fact that
an EDF approach has a strong theoretical underpinning \cite{hk}. The
effective forces used to derive the EDF are nothing else but a
vehicle, since in themselves they have no well--defined physical
meaning. For example, the effective Skyrme two--particle interaction
is neither a particle--hole nor a particle--particle interaction. The
particle--hole interaction (or the Landau parameters) is defined only
as the second order functional derivative of the total EDF with
respect to various densities, while the particle--particle interaction
responsible for the pairing correlations in nuclei has to be supplied
independently and with no logical connection to the Skyrme parameters.

We shall not attempt to even mention various mean--field approaches
suggested so far, in this respect see Refs. \cite{others}, but we
shall concentrate instead on a single aspect of the nuclear EDF,
namely its pairing properties. Only recently it became clear that a
theoretically consistent local EDF formulation of the nuclear pairing
properties is indeed possible \cite{abyy,ab,ablec}. Even though the
crucial role of the pairing phenomena in nuclei has been established
firmly, it is surprising to realize how poor the quality of our
knowledge still is. Phenomenologically, one cannot unambiguously
decide whether the pairing correlations in nuclei have a volume or/and
a surface character
\cite{jacek,dob,dens,volsurf,gogny,goriely,pearson,fayans}.  The
isospin character of the nuclear pairing correlations requires further
clarification as well. These questions become even sharper in the
language of a local EDF.

There is also the largely practical issue of whether one should use a
zero--range or a finite--range effective pairing interaction. The only
reason for the introduction of a finite--range was to resolve the
formal difficulty with divergences in calculating the anomalous
densities \cite{gogny}. The majority of practitioners favor a much
simpler approach, which embodies essentially the same physics, the
introduction of an explicit energy cut--off. The best example are
perhaps the works of the group \cite{goriely,pearson}, which so far is
the leader in describing all known nuclear masses.  An explicit
finite--range of the pairing interaction $r_0$ (which can be
translated into an energy cut--off $E_c\approx \hbar^2/mr_0^2$) and a
zero--range with an explicit energy cut--off $E_c$ in the final
analysis are equivalent. Both approaches, however, are a poor's man
solution to the renormalization problem and reflect simply a lack of
understanding of the role of high--momenta in the pairing channel.
Neither the energy cut--off $E_c$ nor the finite--range of the
interaction $r_0$ carry any physical information and they are simply
means towards getting rid of infinities. The argument that nuclear
forces have a finite--range is superfluous, see
Refs. \cite{abyy,ab,ablec}, since nuclear pairing phenomena are
manifest at small energies and distances of the order of the coherence
length, which is larger than nuclear radii.

We shall consider local nuclear EDFs only (which depend on various
densities, as opposed to an explicit dependence on the full density
matrix), as they proved overwhelmingly successful in describing normal
nuclear properties.  It is natural to expect that the same should
apply to pairing properties.  According to the general theorem of
Hohenberg and Kohn \cite{hk} for many fermion systems there exists a
universal EDF. Unfortunately there are no hints on how to derive such
a functional. In the case of nucleons such a functional should satisfy
some general constraints: rotational invariance, isospin invariance,
time--reversal invariance and conservation of parity. Isospin symmetry
is broken by Coulomb interaction, proton--neutron mass difference and
charge symmetry breaking forces, the last two leading to rather small
effects \cite{fayans,nolen}.  In the case of Coulomb interaction
mainly the direct term has to be accounted for, as the exchange and
correlation Coulomb energies seem to cancel each other to some extent
and their combined effect together with the effect of charge symmetry
breaking forces is relatively small and responsible mainly for such
rather subtle effects as the Nolen--Schiffer anomaly
\cite{fayans,nolen}.  We shall not consider here the contributions due
to Coulomb exchange and charge symmetry breaking energies
\cite{nolen}.  Such a structure of the normal nuclear EDF
${\cal{E}}_N(\bbox{r})$ would be complete in the absence of
superfluidity. Since pairing correlations in nuclear systems are known
to be of rather small amplitude, we shall consider only a superfluid
EDF $ {\cal{E}}_S(\bbox{r})$ of the following structure:
%----------------------------------------------------------------
$$
{\cal{E}}_S(\bbox{r}) = g_0(\bbox{r})|\nu _p(\bbox{r})+\nu _n(\bbox{r})|^2
+g_1(\bbox{r})|\nu _p(\bbox{r})-\nu _n(\bbox{r})|^2 ,
$$
%----------------------------------------------------------------
where $\nu_{p,n}(\bbox{r})$ are the $S=0$ proton/neutron anomalous
densities. There is no firm evidence of pairing in other partial waves
except the BCS--like $s$--wave in either proton or neutron channels
and the evidence for neutron--proton pairing is inconclusive so
far. Notice that ${\cal{E}}_S(\bbox{r})$ is symmetric under the
proton--neutron exchange.  We assume that the effective couplings
$g_{0,1}(\bbox{r})$ might depend on position through the normal
densities and that this dependence is consistent with expected
symmetries.  The density dependence of the effective couplings
$g_{0,1}(\bbox{r})$ arises from two different sources. Firstly, the
bare coupling constant in the pairing channel could in principle have
some intrinsic density dependence and such dependence has been
considered by various authors during the years
\cite{others,jacek,dob,dens,volsurf,fayans}. Secondly, the
renormalization of the pairing interaction, as described in our recent
work \cite{abyy,ab,ablec}, leads to position dependence as well.  The
equations for the quasi--particle wave functions $u_i (\bbox{r})$ and
$v_i (\bbox{r})$ and related quantities are
%====================================================================
\bea
& & E_{gs} = \int d^3r [{\mathcal{E}}_N(\bbox{r})
   +{\mathcal{E}}_S(\bbox{r})],\nn \\
& & {\mathcal{E}}_S(\bbox{r}):= -\Delta(\bbox{r})\nu_c(\bbox{r})
   = g_{\mathit{eff}}(\bbox{r})|\nu_c(\bbox{r})|^2 ,\nn \\
& & \left \{ \begin{array}{l}
 [h (\bbox{r})  - \mu] u_i (\bbox{r})
     + \Delta (\bbox{r})  v_i (\bbox{r})
    = E_i u_i (\bbox{r}) ,  \\
  \Delta^* (\bbox{r}) u_i (\bbox{r})  -
    [ h (\bbox{r}) - \mu ] v_i (\bbox{r})
     = E_i v_i (\bbox{r}),
\end{array}
\right . \nn \\
& & h (\bbox{r})
 = -\bbox{\nabla} \frac{\hbar^2}{2m(\bbox{r})}\bbox{\nabla}
 + U(\bbox{r}),\quad
% & &
\Delta(\bbox{r} )
   := -g_{\mathit{eff}}(\bbox{r})\nu_c(\bbox{r}),
 \nn \\
& & \frac{1}{ g_{\mathit{eff}}(\bbox{r})}=
\frac{1}{g(\bbox{r})} \nn % \\
% & &
 -\frac{m k_c(\bbox{r})}{2\pi^2\hbar ^2}
\left [ 1
  -\frac{k_F(\bbox{r})}{2 k_c(\bbox{r})}
\ln \frac{k_c(\bbox{r})+k_F(\bbox{r})}{k_c(\bbox{r})-k_F(\bbox{r}) }
    \right ]    \nn \\
& & \rho_c(\bbox{r}) =
     \sum _{ E_i\ge 0}^{ E_c} 2|v_i(\bbox{r})|^2,\quad
    \nu_c(\bbox{r}) =
     \sum _{ E_i\ge 0}^{ E_c} v_i^*(\bbox{r})u_i(\bbox{r}),   \nn \\
& &    E_c  +\mu =
 \frac{\hbar^2k_c^2(\bbox{r})}{2m(\bbox{r})} + U(\bbox{r}),
\;           \mu =
 \frac{\hbar^2k_F^2(\bbox{r})}{2m(\bbox{r})} + U(\bbox{r}). \nn
\eea
%====================================================================
For the sake of simplicity we do not display the spin and isospin
variables. $k_F(\bbox{r})$ is the local Fermi momentum, which could be
either real or imaginary, while $k_c(\bbox{r})$ is real
\cite{abyy,ab,ablec}. The role of the particle continuum
\cite{jacek,ab0} is taken into account exactly using the technique
described in Refs. \cite{stb}, the contour integration in the
complex energy plane of the Gorkov propagators for the Bogoliubov
quasi--particles, in order to evaluate various densities. All
calculations have been performed in coordinate representation and all
nuclei have been treated as spherical.  For reasons we discussed in
detail in Refs. \cite{abyy,ab,ablec}, the cut--off energy should be
chosen of the order $E_c={\cal{O}}(\eps_F)$. In practice we found that
a value $E_c\approx 70$ MeV for SLy4 \cite{sly} and $E_c\approx 55$
MeV for FaNDF$^0$ \cite{fayans} is satisfactory and it ensures a
convergence of the pairing field $\Delta(\bbox{r})$ with a relative
error of $\approx 10^{-5}$ for density independent bare couplings.
Note that the calculation of $\Delta(\bbox{r})$ alone would require a
significantly smaller $E_c$ of order 10--15 MeV \cite{ab,ablec}. The
optimal value for $E_c$ varies, depending on whether one uses an
effective mass close to the bare nucleon mass or a reduced one, as is
typical with Skyrme interactions.  Even though this explicit cut--off
energy $E_c$ appears in various places, indeed, no observable shows
any dependence on $E_c$, when its value it is chosen
appropriately. Upon renormalization of the zero--range pairing
interaction, the emerging formalism is no more complicated than a
simple energy cut--off approach, with the only major bonus however,
that there is no energy cut--off dependence of the results.  Since the
kinetic energy of the system is a diverging quantity of $E_c$ and only
the total energy is a convergent quantity \cite{abyy,ab,ablec,george}
it is very important that all densities (normal and anomalous) be
evaluated using the same energy cut--off $E_c$.

We shall treat even and odd number of particles within the same
framework and using the same EDF parameterization, unlike e.g.
Refs. \cite{goriely,pearson}. The formalism for evaluating the Gorkov
propagators for odd systems is described in great detail in
Refs. \cite{fayans,migdal}. For the normal part of the EDF we shall
use either the Lyon parameterization of the Skyrme interaction
\cite{sly} and or the FaNDF$^0$ suggested by Fayans
\cite{fayans}. Both EDFs reproduce with high accuracy the infinite
matter equations of state of Refs. \cite{nm}.

%---------------------------------------------------------------------
%---------------------------------------------------------------------
\begin{figure}[tbh]

\begin{center}

\epsfxsize=7.0cm

\centerline{\epsffile{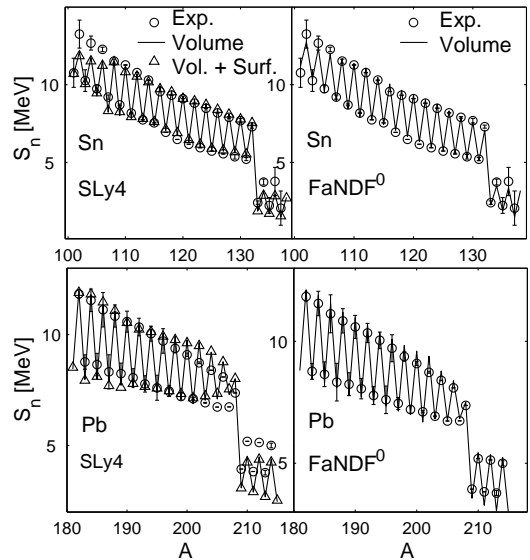}}

\end{center}

\caption{$S_n$  for tin and lead
isotopes computed using the SLy4 EDF (left) and Fayans' FaNDF$^0$
(right) with either volume or half volume--half surface pairing.
}
\label{fig:fig2}

\end{figure}
%---------------------------------------------------------------------
%---------------------------------------------------------------------

We shall present here results only for those nuclei for which we can
make comparison with available recommended nuclear masses \cite{audi}.
We consider at first the tin (38 nuclei) and lead (34 nuclei) isotope
chains.  We performed a number of calculations of these isotopes
essentially from the neutron to the proton drip lines, see
Ref. \cite{ablec} for some preliminary results.  For these nuclei we
can test only the sum of the coupling constants,
namely $g(\bbox{r})= g_0(\bbox{r}) +
g_1(\bbox{r})$. We have considered a bare coupling
$g(\bbox{r})=const$, which corresponds to volume pairing, and also
$g(\bbox{r})=V_0[1-\rho(\bbox{r})/\rho_c]$, with parameters chosen to
describe roughly one half volume and one half surface pairing, as
suggested in particular in Ref. \cite{dob}.  One--neutron separation
energies $S_n$ and two--neutron separation energies $S_{2n}$ for tin
and lead isotopes were computed for constant pairing
$g(\bbox{r})=const$, with mean--field computed with either SLy4
interaction \cite{sly} ($g(\bbox{r})=-250 \; {\mathrm{MeV}} \;
{\mathrm{fm}}^{3}$) or with Fayans normal nuclear EDF \cite{fayans}
($g(\bbox{r})=-200 {\mathrm{MeV}}\;{\mathrm{fm}}^{3}$).  For the case
of SLy4 interaction we also show results obtained for the half
volume--half surface pairing model ($ V_0=-370 \; {\mathrm{MeV}} \;
{\mathrm{fm}} ^{3}, \rho_c = 0.32 \; {\mathrm{fm}} ^ {-3} $). The
search for the appropriate values for $g(\bbox{r})$ was performed only
among a finite set of values, e.g. in the case of volume pairing we
considered $g=-200, -225, -250, -275$ and -300 ${\mathrm{MeV}} \;
{\mathrm{fm}} ^{3}$, see also Ref. \cite{ablec}.

%---------------------------------------------------------------------
%---------------------------------------------------------------------
\begin{figure}[tbh]

\begin{center}

\epsfxsize=7.0cm

\centerline{\epsffile{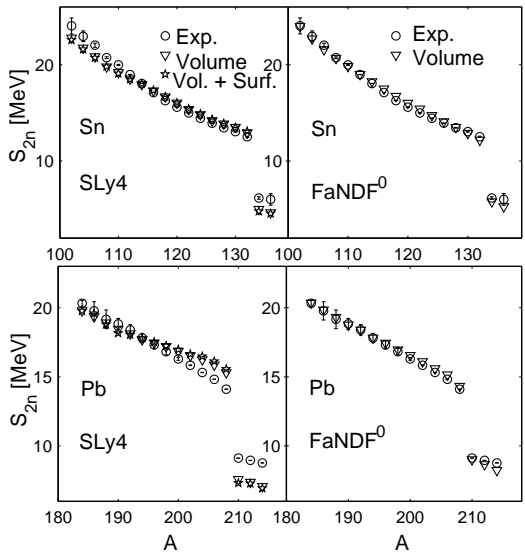}}

\end{center}

\caption{The same as in Fig. \ref{fig:fig2} but $S_{2n}$.}

\label{fig:fig3}

\end{figure}
%---------------------------------------------------------------------
%---------------------------------------------------------------------

%---------------------------------------------------------------------
%---------------------------------------------------------------------
\begin{figure}[tbh]
\begin{center}
\epsfxsize=7.0cm
\centerline{\epsffile{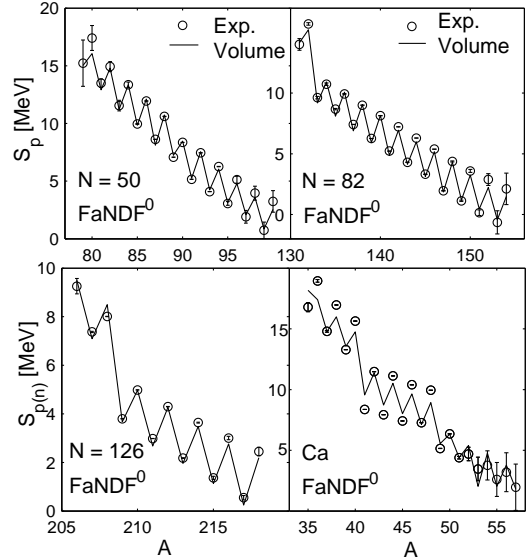}}
\end{center}
\caption{ $S_p$ for isotone chains $N=$ 50, 82 and 126 and $S_n$ for
calcium isotopes.  For $S_{2p}$ and $S_{2n}$ see
online report [22]. }
\label{fig:fig4}
\end{figure}
%---------------------------------------------------------------------
%---------------------------------------------------------------------

%============================================================================
\noindent{Table 1. Rms of $S_{2N}$ and $S_N$ deviations respectively from
experiment \cite{audi} (in MeV's) for several isotope and isotone chains.}

\begin{tabular}{||l|l|l|l||} \hline \hline
Z or N    &  $S_{2N}/S_N$ &  $S_{2N}/S_N$     &  $S_{2N}$              \\
chain     &  present     & Ref. \cite{goriely} &  Ref. \cite{meng}  \\
\hline \hline
Z =  20  & 0.82/0.76    & 1.02/0.92   & 0.96             \\ \hline
Z =  28  & 0.67/0.50    & 0.66/0.55   & 1.30             \\ \hline
Z =  40  & 0.93/0.63    & 0.66/0.63   & 2.21             \\ \hline
Z =  50  & 0.29/0.21    & 0.43/0.35   & 0.95             \\ \hline
Z =  82  & 0.23/0.37    & 0.58/0.53   & 0.74             \\ \hline
N =  50  & 0.37/0.26    & 0.41/0.23   & NA               \\ \hline
N =  82  & 0.43/0.31    & 0.50/0.56   & NA               \\ \hline
N = 126  & 0.42/0.23    & 0.88/0.52   & NA               \\ \hline \hline
\end{tabular}
%============================================================================

The agreement between experiment and theory is particularly good for
the case of FaNDF$^0$. We relate this result with the fact that the
effective mass in FaNDF$^0$ is the bare nucleon mass, unlike the case
of SLy4 EDF and in agreement with the global mass fit of
Ref. \cite{goriely,pearson}. It is notable that the agreement with experiment
is equally good for both tin and lead isotopes with the same value of
$g$, unlike Refs. \cite{fayans}.  Even though we used the same
normal nuclear EDF as in Refs. \cite{fayans}, our agreement
with experiment is notable superior, see Ref. \cite{report},
even though we parameterize the pairing interaction with one parameter
only vs. up to five parameters used in these papers.

Since FaNDF$^0$ in conjuction with the bare pairing coupling constant
$g(\bbox{r}) =-200 \; {\mathrm{MeV}} \;{\mathrm{fm}} ^{3}$ apparently
provides the best description in case of tin and lead isotopes,
further calculations were performed only with this choice of
parameters. In Fig. \ref{fig:fig4} we display the one--proton
separation energies $S_p$ for three isotone chains (23 nuclei with
$N=50$, 25 nuclei with $N=82$ and 14 nuclei with $N=126$) and $S_n$
for calcium isotopes (24 nuclei), and for nickel and zirconium
isotopes see Ref. \cite{report}, 212 nuclei in total.  Since the
neutron numbers for these isotone chains are also magic, again, we can
test only the same combination of coupling constants $g(\bbox{r})=
g_0(\bbox{r}) + g_1(\bbox{r})$. As we have conjectured at the
beginning of this study, one can indeed describe with a single value
$g$ separately for proton and neutron pairing correlations in both even
and odd systems, as opposed to the treatment of
Refs. \cite{goriely,pearson}, which slightly violates isospin
invariance.  In essentially all cases in which we have been able to
perform a comparison between our results and those available in
literature, our results were either qualitatively superior or, in a
few separate cases, as good as any other results. In Table 1 we
present rms deviation from experimental (recommended) values
\cite{audi} for the two and one nucleon separation energies for
several isotope and isotone chains. The size of each set of nuclei in
a chain was given by the number of nuclei in Ref. \cite{goriely}, for
which there are experimental values in the unpublished Audi and
Wapstra 2001 compilation.

There are a number of theoretical arguments, suggesting that the
pairing coupling should be density/position dependent, due to the
coupling to surface/particle--hole modes, e.g. Ref. \cite{terasaki}.
A similar line of reasoning was presented in the case of dilute
systems \cite{gorkov,hh} and neutron matter \cite{clark,nstar} for
quite some time. Our results, see
Figs. \ref{fig:fig2}--\ref{fig:fig3}, show that $S_n$ and $S_{2n}$ for
tin and lead isotopes are not particularly sensitive to such effects.
To some extent this is not a surprise, since pairing correlations are
''built'' at distances of the order of the coherence length $\xi
\propto \hbar^2 k_F/m\Delta$ \cite{degennes}, see Ref. \cite{vortex}
for a related instructive example. This apparent low sensitivity of
$S_N$ and $S_{2N}$ to a possible density dependence of the pairing
couplings, could in principle be profitably used to describe other
observables. From the results of Refs. \cite{hh} one might infer that
pairing coupling constants could have a noticeable variation with the
isospin composition of a given system, since the magnitude of the
induced interactions changes dramatically as the number of fermion
species varies. Neither our results, nor previous work has
necessitated the introduction of such a dependence however. In our
phenomenological approach, based on general symmetry arguments alone
and the fact that the pairing correlations are relatively weak in
nuclear systems, we restricted the form of the EDF superfluid
contribution to the simplest one compatible with known symmetries.  We
were able to infer that pairing properties of either kind of nucleons
can be accounted for with a single constant $g=g_0+g_1$. It remains to
be seen whether the other (non--perturbative) combination
$g^\prime=g_0-g_1$ (never considered by other authors) could ever
become relevant.

%--------------------------------------------------------------------
%--------------------------------------------------------------------
%\newpage

\end{document}